# Lyman α Emission from High-redshift Galaxies

Max Pettini[1], Richard W. Hunstead[2], David L. King[1] and Linda J. Smith[3]

[1] Royal Greenwich Observatory, Madingley Road, Cambridge, CB3 0EZ, UK
[2] School of Physics, University of Sydney, NSW 2006, Australia
[3] Department of Physics and Astronomy, University College London, Gower Street, London WC1E 6BT, UK

**Abstract.** We summarise the results of a deep search for Lyman α emission from star-forming regions associated with damped Lyman α absorption systems and conclude that the Lyman α luminosity of high redshift galaxies is generally less than $10^{42}$ erg s$^{-1}$. We also present a newly discovered case, in the field of the QSO Q2059−360, where the emission is unusually strong, possibly because the damped system is close in redshift to the QSO.

Damped Lyman α systems at $z \simeq 2-3$ arise in galaxies which are generally metal-poor and, in some cases, may have experienced only a few episodes of star formation (Pettini et al. 1995). It is therefore reasonable to expect that the fields of QSOs with damped systems may be good candidates in searches for Lyman α emission from high-redshift H II regions, particularly as the QSO light is completely extinguished by the damped Lyman α absorption line over a redshift interval which can span up to several tens of Å.

Over the last few years we have conducted such a search with high-resolution slit spectroscopy of 21 damped systems in 18 QSOs. Two noteworthy features of our approach are the use of photon-counting detectors, which maintain the photon statistics even at the very low count rates measured in the cores of the damped lines (essentially just the dark sky background), and the long integration times (between $\sim 10\,000$ and $\sim 40\,000$ s). Consequently, we are able to reach some of the lowest flux limits reported, corresponding to Lyman α luminosities $L_\alpha(3\sigma) \leq 1 \times 10^{42}$ erg s$^{-1}$ ($H_0 = 50$ km s$^{-1}$ Mpc$^{-1}$, $q_0 = 0.5$).

The principal result is that high-redshift galaxies are *not* strong Lyman α emitters; Lyman α emission is below, or close to, the detection limit in nearly all cases surveyed. We have found only two definite detections, associated with the $z_{\rm abs} = 2.4651$ system in Q0836+113 – reported by Hunstead, Pettini and Fletcher (1990) and confirmed by new observations obtained in 1994 – and with the $z_{\rm abs} = 3.0831$ system in Q2059−360 – shown in Fig.1. In addition, we have two or three cases of marginal detections.

From Fig.1 it is evident that there is excess signal in the base of the damped line, which we interpret as Lyman α emission redshifted by $\sim 470$ km s$^{-1}$ from the absorption system. The integrated flux is $F_\alpha = 7 \times 10^{-17}$ erg cm$^{-2}$ s$^{-1}$, corresponding to a Lyman α luminosity $L_\alpha =$ $5 \times 10^{42}$ erg s$^{-1}$. At the relatively coarse spatial resolution of our observations the emission appears to be approximately coincident with the QSO position. However, the velocity difference between emission and absorption suggests that the Lyman $\alpha$ emitting region is not associated directly with the damped system, but more likely arises in a galaxy which is a member of a cluster also including the absorber.

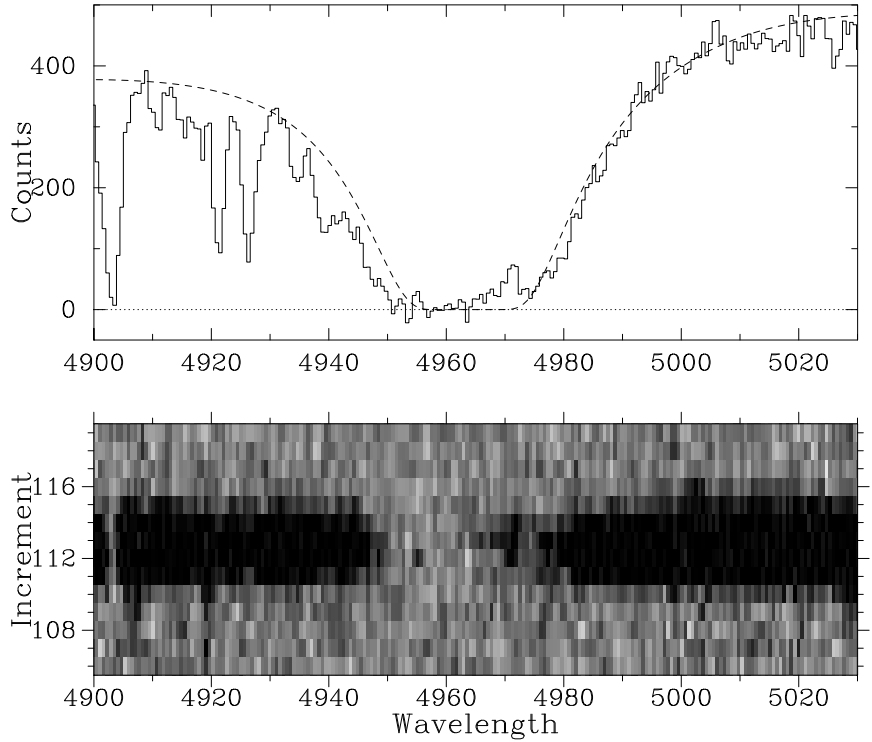

**Fig. 1.** *Top*: Observed profile of the $z_{\rm abs} = 3.0831$ damped Lyman $\alpha$ line in the $z_{\rm em} = 3.13$ QSO Q2059−360; the x-axis gives the wavelength in Å. The broken line shows the theoretical profile for a neutral hydrogen column density $N({\rm H}^0) = 5 \times 10^{20}$ cm$^{-2}$, centred at the redshift of Lyman $\beta$ and metal absorption lines in the same system. *Bottom*: Grey-scale ($\pm 3\sigma$ of the sky signal) reproduction of the corresponding portion of the two-dimensional spectrum; each spatial increment is 0.75 arcsec, or $\sim 5$ kpc at $z = 3.08$. The spectrum was recorded with 8000 s integration using a Tektronix blue-sensitive CCD at the cassegrain focus of the Anglo-Australian telescope. The resolution is 1.4 Å *FWHM*; the data were obtained over two nights in September 1994, when the seeing was $\sim 1 - 1.5$ arcsec *FWHM*.

In this respect Q2059−360 is a similar case to the field of Q0528−250 where Moller and Warren (1993) found emission at $z_{em} = 2.811$. It may not be a coincidence that two out of the only three known detections of Lyman α emission in the core of a damped Lyman α line are absorption systems close to the QSO redshift. It remains to be seen whether these are environments favouring large bursts of star formation, or whether the Lyman continuum radiation from the QSO itself is responsible for the unusually strong emission feature.

The lack of significant Lyman α emission from damped Lyman α galaxies is in line with the generally null results of other Lyman α searches (e.g. Pritchet 1994) and has motivated a considerable amount of theoretical work. The observations confirm current ideas that: *(i)* even massive bursts of star formation are probably bright in Lyman α for only a relatively brief period (Charlot and Fall 1993); *(ii)* Lyman α line radiation can be quenched very effectively by small amounts of dust, due to the large optical depths involved (Chen and Neufeld 1994); and *(iii)* Lyman α photons in any case probably escape from the parent galaxy along preferred directions determined by the inhomogeneous nature of the interstellar medium (Neufeld 1991).

The continuing improvement in infrared detectors and the encouraging results of pilot observations (Hu et al. 1993; Bunker et al. 1995) suggest that, in future, searching for Balmer lines at $z > 2$ may be the most promising strategy for assessing the relative importance of these processes and measuring the star-formation rate of high-redshift galaxies.